\documentclass[final,3p,times,twocolumn,fixfloat]{elsarticle}
\usepackage{graphicx}
\usepackage{amssymb} 
\usepackage{amsmath} 

\usepackage{lineno}

 \usepackage{ifpdf}
 \ifpdf
 \usepackage[pdftex]{hyperref}
 \else
 \usepackage[hypertex]{hyperref}
 \fi

 \hypersetup{
   pdftitle={},%
   pdfauthor={},%
   pdfsubject={},%
   pdfkeywords={},%
   pdfstartview={},%
   bookmarksopen=true, breaklinks=true, debug=true, %
   colorlinks=true, linkcolor=red, citecolor=blue, urlcolor=blue
 }

\journal{Physics Letters B}

\usepackage{float} 
\usepackage{subfig} 

\usepackage{amsmath}
\usepackage{epsfig}
\usepackage{float}
\usepackage{verbatim} 
\usepackage{color} 

\newcommand{\ep}{\varepsilon}

\newcommand{\eqs}[1]{\begin{equation} \begin{split} #1\end{split} \end{equation} }

\newcommand{\ie}{{\it i.e.}}
\newcommand{\eg}{{\it e.g.}}

\newcommand{\Q}{{\cal Q}}

\newcommand{\cf}[1]{{Fig.~\ref{#1}}}

\begin{document} 
\begin{frontmatter}

\title{Reassessing the importance of the colour-singlet contributions to direct $J/\psi+W$ production at the LHC and the Tevatron}

\author[IPNO]{J.P.~Lansberg}
\author[IPNO,LPT]{C. Lorc\'e}
\address[IPNO]{IPNO, Universit\'e Paris-Sud, CNRS/IN2P3, F-91406, Orsay, France}
\address[LPT]{LPT, Universit\'e Paris-Sud, CNRS, F-91405, Orsay France}

\begin{abstract}
\small
We show that the colour-singlet contributions to the hadroproduction of $J/\psi$ in association with a $W$ boson 
are sizable, if not dominant over the colour-octet contributions. 
They are of two kinds, $sg\to J/\psi + c +W$ at $\alpha^3_S \alpha$ and $q\bar q' \to \gamma^\star\!/Z^\star W\to J/\psi W$
 at order $\alpha^3$. These have not been considered in the literature until now. Our conclusion is that the hadroproduction of a $J/\psi$ 
in association with a $W$ boson cannot be claimed as a clean probe of the colour-octet mechanism. The rate 
are small even at the LHC and it will be very delicate to disentangle the colour-octet contributions from
the sizable colour-singlet ones and from the possibly large double-parton-scattering contributions. During this analysis, 
we have also noted that, for reactions such as the production of a $J/\psi$ by light quark--antiquark fusion, the 
colour-singlet contribution via an off-shell photon is of the order of the expectation from the colour-octet 
contribution via an off-shell gluon. This is relevant for inclusive production at low energies close to the 
threshold. Such an observation also likely extends to other processes naturally involving light-quark annihilation.
\end{abstract}

\begin{keyword}
\small
  Quarkonium\ production \sep Vector boson production 
\PACS  12.38.Bx \sep 14.40.Gx \sep 13.85.Ni
\end{keyword}

\end{frontmatter}




\section{Introduction}

Since the mid eighties, the field of quarkonium physics has faced a number of puzzles challenging our understanding of QCD at the interplay between its short- and long-distance domains. The puzzles related to the quarkonium production at the Tevatron have been attributed to the colour-octet mechanism (COM), \emph{i.e.} the non-perturbative transition of heavy quark-antiquark pairs in colour-octet state into quarkonia (see~\cite{Lansberg:2006dh,Brambilla:2010cs,ConesadelValle:2011fw} for reviews). 

Since a few years, we know that $\alpha^4_S$ and $\alpha^5_S$ corrections to the colour-singlet mechanism (CSM)~\cite{CSM_hadron} are essential to try to explain the $P_T$ dependence of the $J/\psi$ and $\Upsilon$ cross sections observed in high-energy hadron collisions~\cite{Campbell:2007ws,Artoisenet:2007xi,Gong:2008sn,Gong:2008hk,Artoisenet:2008fc,Lansberg:2008gk}. Polarisation predictions are also dramatically affected by QCD corrections, both in the inclusive case and in the production of quarkonia with a prompt photon~\cite{Gong:2008sn,Gong:2008hk,Artoisenet:2008fc,Lansberg:2010vq,Li:2008ym,Lansberg:2009db}. As far as the $P_T$-integrated yield is concerned, colour-singlet $Q \bar Q$ configurations have been shown 
to be sufficient\footnote{The CSM is nonetheless known to be plagued by infrared divergences in the case of $P$-wave decay at NLO, 
earlier regulated by an ad-hoc binding energy~\cite{Barbieri:1976fp},  
 which can however
be rigorously cured~\cite{Bodwin:1992ye} in the more general framework of NRQCD.} to account for the experimental data~\cite{Brodsky:2009cf,Lansberg:2010cn}.

\begin{figure}[h!]
\centering
\subfloat[]{\includegraphics[scale=.375,draft=false]{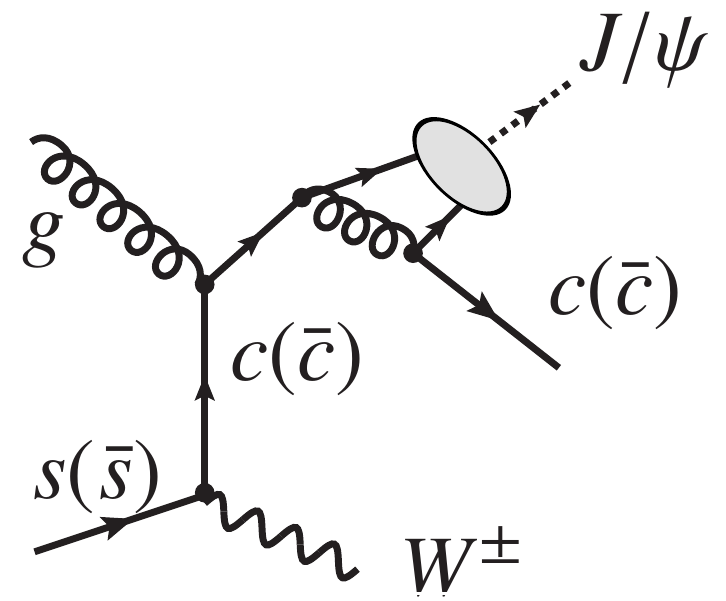}\label{diagram-a}}\hspace*{-.2cm}
\subfloat[]{\includegraphics[scale=.375,draft=false]{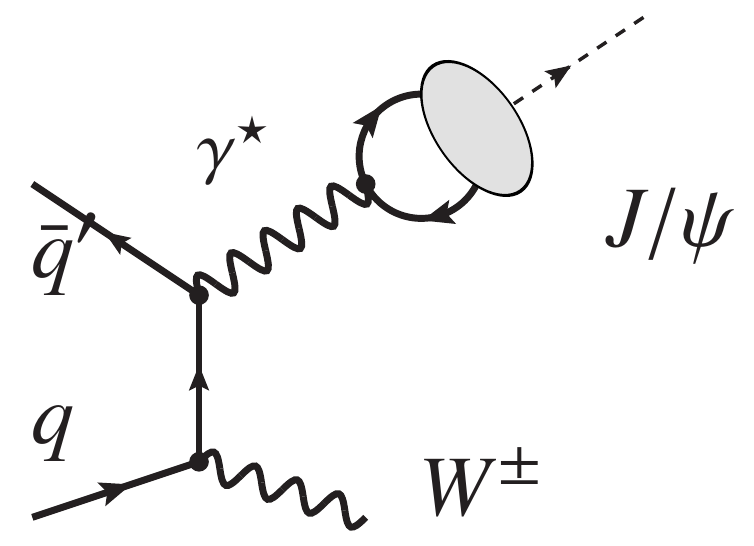}\label{diagram-b}}\hspace*{-.2cm}
\subfloat[]{\includegraphics[scale=.375,draft=false]{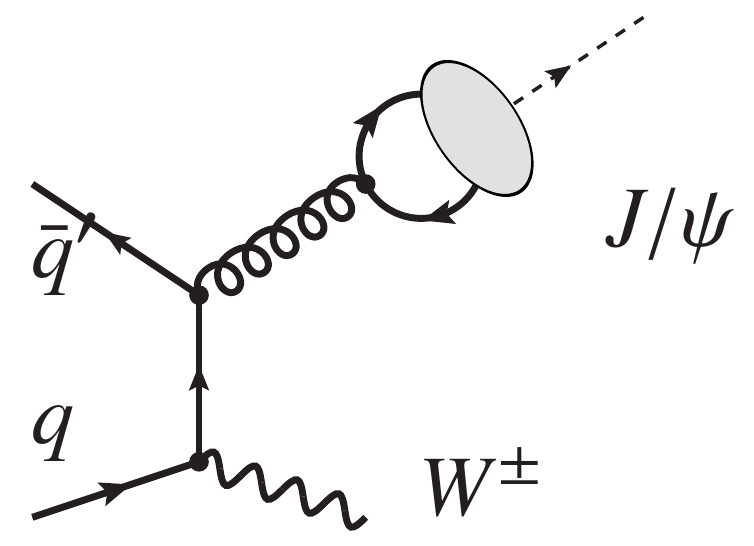}\label{diagram-c}}\hspace*{-.2cm}
\caption{Representative diagrams contributing to $J/\psi+ W^\pm$ hadroproduction in the CSM at orders $\alpha_S^3\alpha$ (a), $\alpha^3$ (b) and in the COM at orders $\alpha_S^2\alpha$ (c). The quark and antiquark attached to the ellipsis are taken as on-shell and their relative velocity $v$ is set to zero.}
\label{diagrams}
\end{figure}

In this Letter, we reassess the importance 
of the leading-$v^2$ contribution to $J/\psi+W^\pm$ -- \ie~ from the colour-singlet transitions. 
In the previous analyses of $J/\psi+W$~\cite{Barger:1995vx,Kniehl:2002wd,Li:2010hc}, these have been disregarded since
formally appearing at higher orders in $\alpha$ or $\alpha_S$. In fact, they are not negligible at all.
Quotes such as {\it ``$\psi+W$ offers a clean test of the colour-octet contributions''} from~\cite{Barger:1995vx} 
and {\it ``If the $J/\psi+W$ production is really detected, it would be a solid basis for testing the color-octet 
mechanism of the NRQCD''} from~\cite{Li:2010hc} are overstated if not misleading. This observable is not cleaner
than the inclusive production for instance.

We have identified two classes of important colour-singlet contributions. The first comes from the strange-quark--gluon fusion which produces a $W+c$  pair where the charm quark fragments into a $J/\psi$ (see \cf{diagram-a}). This is reminiscent of the leading-$P_T$ contribution to $J/\psi + c\bar c$ for instance~\cite{Artoisenet:2007xi}. In the past, $W+c$ has indeed been identified as a probe of the strange quark PDF~\cite{Baur:1993zd}\footnote{Pending the available statistics, $J/\psi + W+c$, as $W+c$, could in principle be discriminated experimentally owing to the presence of an additional charmed hadron in the final state.}. The other class is simply a contribution {\it \`a la} vector-meson dominance. The $^3S_1$ quarkonium bound-state is simply produced by an off-shell photon (or $Z$) emitted by the quark which also radiates the $W$ boson (see \cf{diagram-b}). 

The latter contribution is clearly enhanced in $p \bar p$ collisions at the Tevatron owing to the presence of valence antiquarks in the antiproton, whereas the former contribution is getting larger at LHC energies in $pp$ collisions thanks to the enhancement of the gluon PDF at lower $x$. In any case, these CSM processes are not at all negligible compared to the leading colour-octet contributions (see \cf{diagram-c}). Interestingly, both these Born contributions possess a leading-$P_T$ contribution ($P_T^{-4}$). Not only are they significant at low $P_T$, but they remain large at large $P_T$. This is at variance with the inclusive case where the Born contributions are not leading power in $P_T$.

In section 2, we briefly discuss how we have evaluated the cross sections of the different contributions. In section 3, we present and discuss our results. Section 4 gathers a detailed discussion of the relative size of the CSM contribution via an off-shell {\it photon} w.r.t that of the COM via an off-shell {\it gluon}. We finally conclude in section 5.

\section{Cross-section  evaluation }
\label{sec:xsection}

In the CSM~\cite{CSM_hadron}, the amplitude for the production of a $^3S_1$ quarkonium ${\Q}$ of a given momentum $P$ and of polarisation $\lambda$ accompanied by other partons, noted $j$, and a $W$ boson is written as the product of the amplitude to create the corresponding heavy-quark pair, a spin projector $N(\lambda| s_1,s_2)$ and $R(0)$, the radial wave function at the origin in the configuration space. Precisely, one has
\eqs{ \label{CSMderiv3}
{\cal M}&(ab \to {\Q}^\lambda(P)+W+j)=\!\sum_{s_1,s_2,i,i'}\!\!\frac{N(\lambda| s_1,s_2)}{ \sqrt{m_Q}} \frac{\delta^{ii'}}{\sqrt{N_c}} \frac{R(0)}{\sqrt{4 \pi}}\\\times&
{\cal M}(ab \to Q^{s_1}_i \bar Q^{s_2}_{i'}(\mathbf{p}=\mathbf{0}) + W + j),
}
where one defines $P=p_Q+p_{\bar Q}$, $p=(p_Q-p_{\bar Q})/2$, and where $s_1$,$s_2$ are the heavy-quark spin components and $\delta^{ii'}/\sqrt{N_c}$ is the projector onto a colour-singlet state. $N(\lambda| s_1,s_2)$ has a simple expression in the non-relativistic limit: $\frac{\ep^\lambda_{\mu} }{2 \sqrt{2} m_Q } \bar{v} (\frac{\mathbf{P}}{2},s_2) \gamma^\mu u (\frac{\mathbf{P}}{2},s_1) \,\, $ where $\ep^\lambda_{\mu}$ is the quarkonium polarisation vector. Once one sums over the heavy-quark spin components, one obtains traces which can be evaluated in a standard way. In particular, for LO evaluations --without loops-- one can simply use the framework described in~\cite{Artoisenet:2007qm} based on the tree-level matrix element generator {\small MADONIA}~\cite{Madonia}. Another possibility would be to use HELAC-Onia~\cite{Shao:2012iz}.

For the cross-section evaluation, we have used the parameters $|R_{J/\psi}(0)|^2=1.01$ GeV$^3$ and Br$(J/\psi \to \ell^+\ell^-)=0.0594$. Neglecting relativistic corrections, one has in the CSM, $M_{J/\psi}=2m_c$. We have taken  $m_W=80.39$~GeV and $\sin^2(\theta_W)=0.23116$. The uncertainty bands for the resulting predictions are obtained from the {\it combined} variations of the heavy-quark mass within the range $m_c=1.5\pm 0.1$ GeV, with the factorisation $\mu_F$ and the renormalisation $\mu_R$ scales chosen among the couples $((0.75,0.75);(0.75,2);(1,2);(1,1);(2,1);(2,0.75);(2,2))\times m_W$.

For the colour-octet contributions, the only relevant parameter is the NRQCD Long Distance Matrix Elements (LDME) $\langle O_{J/\psi}(^3S_1^{[8]}) \rangle $. We have set it to $2.2 \times 10^{-3}$ GeV$^{3}$, \ie~the value obtained in the recent global NLO analysis of Butenschoen and Kniehl~\cite{Butenschoen:2011yh}. This value is also of the order of what was obtained in another recent NLO NRQCD fit~\cite{Chao:2012iv}.

There are of course drawbacks in using LDME obtained from NLO fits. First, NLO corrections to the hard part of color-octet processes to inclusive production show a $K$ factor higher than one which leads to a reduction of the CO LDME compared to those extracted from a LO fit. Yet, a comparison to the NLO results of~\cite{Li:2010hc} for CO channels indicate that our evaluation is reasonable. Moreover, various CO contributions can interfere and fits can yield negative values. For instance, a recent NLO fit has obtained such a negative result for this LDME~\cite{Gong:2012ug}. It would not make much sense to use such a value in LO computations since the cross section would then be negative. It is therefore important to recall that our choice is also close to the LO analysis of~\cite{Braaten:1999qk} and from analyses which partially took into account QCD corrections \cite{Kniehl:1998qy,Gong:2008ft}.

\section{Results}

Our leading-order results for the differential cross sections in $P_T$ are shown in \cf{fig:dsigdPT} for the Tevatron $(a)$, and for the LHC at $8$ TeV $(b)$ and $14$ TeV $(c)$.

\begin{figure}[hbt!]
\begin{center}
\subfloat[1.96 TeV]{\includegraphics[width=0.8\columnwidth,draft=false]{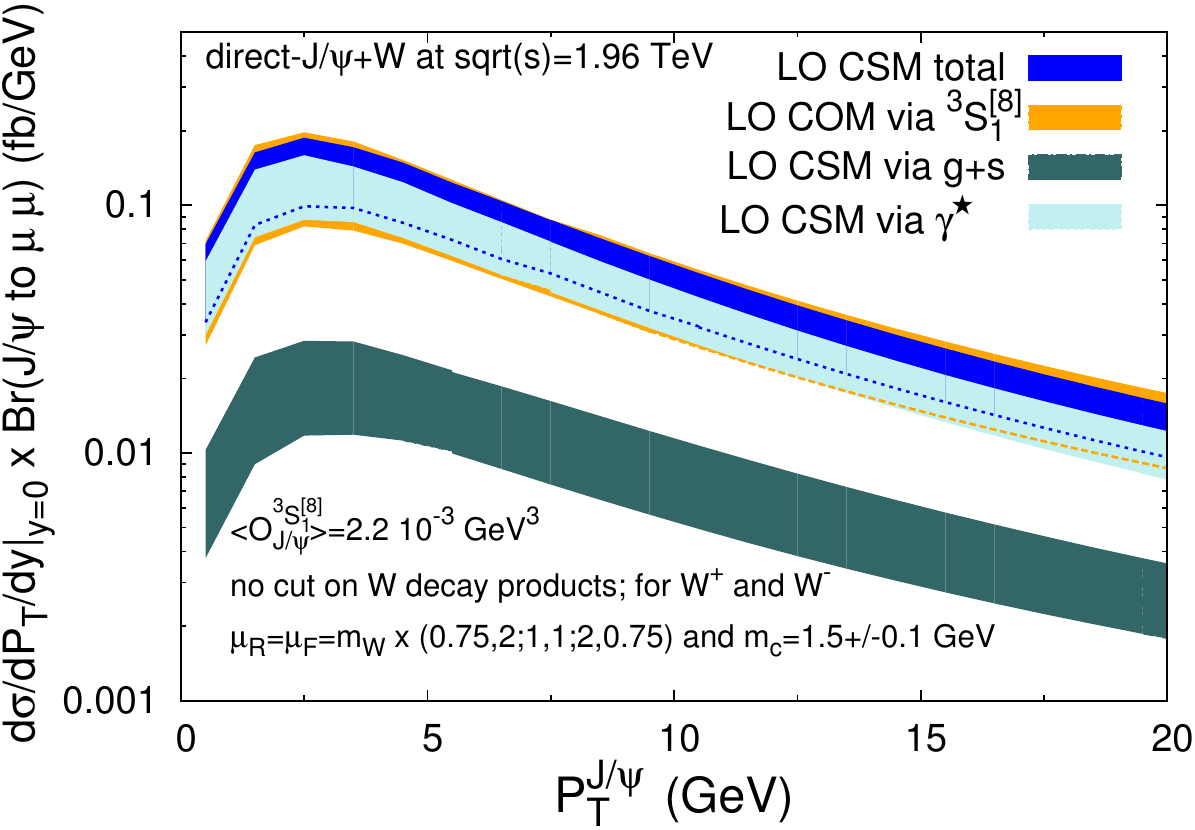}\label{fig:dsigdPTa}}\\\vspace*{-0.25cm}
\subfloat[8 TeV]{\includegraphics[width=0.8\columnwidth,draft=false]{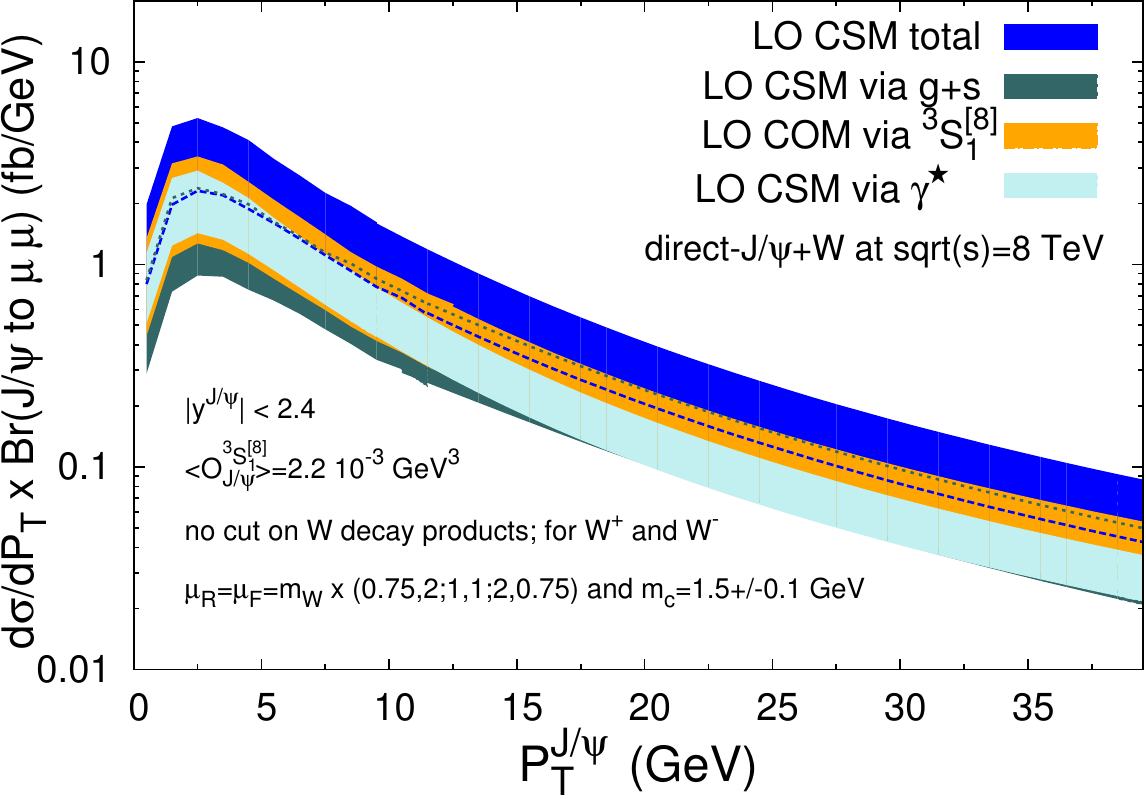}\label{fig:dsigdPTb}}\\\vspace*{-0.25cm}
\subfloat[14 TeV]{\includegraphics[width=0.8\columnwidth,draft=false]{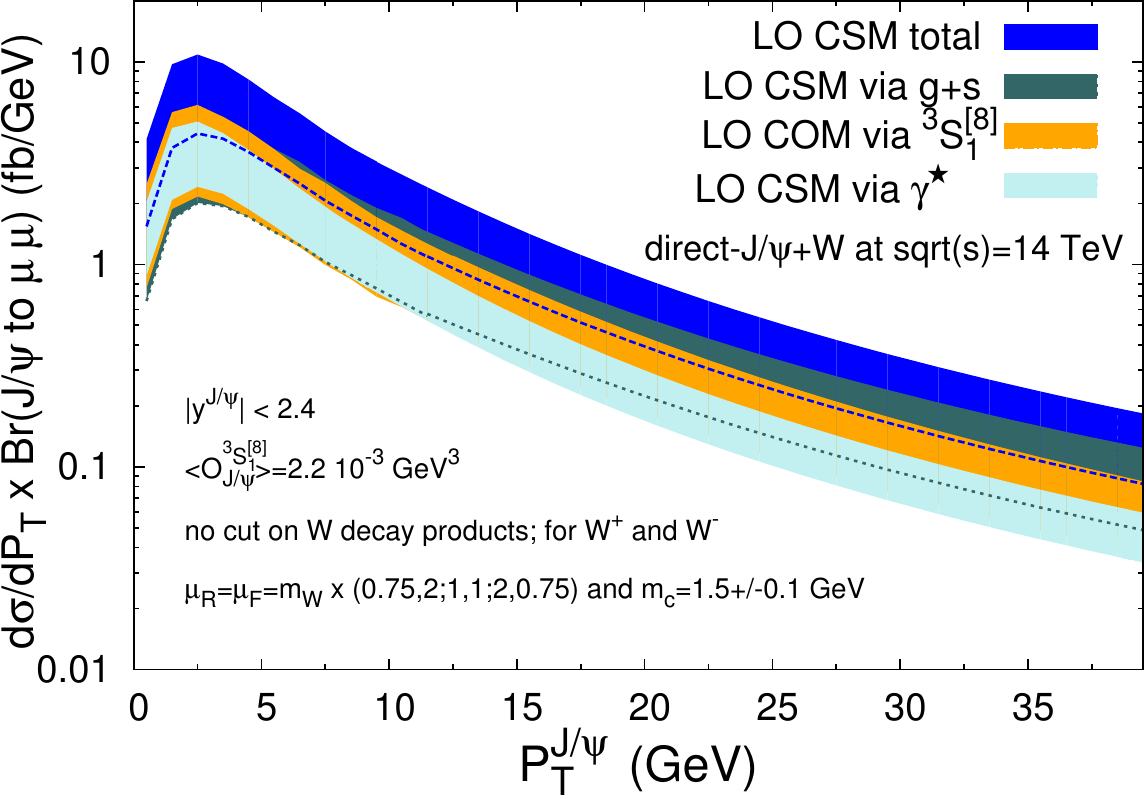}\label{fig:dsigdPTc}}
\caption{Differential cross section at LO for $J/\psi+W$ vs. $P_T$ for the Tevatron $(a)$ and the LHC at $8$ TeV $(b)$ and $14$ TeV $(c)$. The orange band is for the COM while the light blue, dark green and blue bands are for the CSM via $\gamma^\star$, via $sg$ fusion and total contributions, respectively.}
\label{fig:dsigdPT}
\end{center}
\end{figure}

At the Tevatron, the COM contribution\footnote{We note once again that our COM results are compatible with those of~\cite{Kniehl:2002wd} and the LO of~\cite{Li:2010hc} once the differences in the choices of the scales, of the LDME and of the kinematical cuts are taken into account.} (orange band) is significantly larger than that of the CSM via $sg$ fusion (dark green band). However, it is of similar size as the CSM contribution via $\gamma^\star$ (light blue band). Note that the light-blue band actually also contains other electroweak contributions appearing at the same order, \emph{i.e.}~via $Z^\star$, but the yield is strongly dominated by processes via $\gamma^\star$. At LHC energies, the three contributions are of the same order. The total CSM cross section is thus about twice as large as the COM one, probably a bit more at 14 TeV and at large $P_T$ (see \cf{fig:dsigdPTc}).

Such results clearly demonstrate that, contrary to earlier claims in the literature~\cite{Barger:1995vx,Li:2010hc}, the yield for the production of $J/\psi$ in association with a $W$ boson \emph{cannot} actually be used as a clean probe of the COM. This remains true over the whole range in $P_T$. The Born CSM contributions considered here are indeed leading $P_T$ at variance to the inclusive case where leading-$P_T$  contributions~\cite{Artoisenet:2007xi,Artoisenet:2008fc} only appear at higher orders in $\alpha_S$.

\begin{figure}[htb!]
\centering
\subfloat[8 TeV]{\includegraphics[width=.8\columnwidth,draft=false]{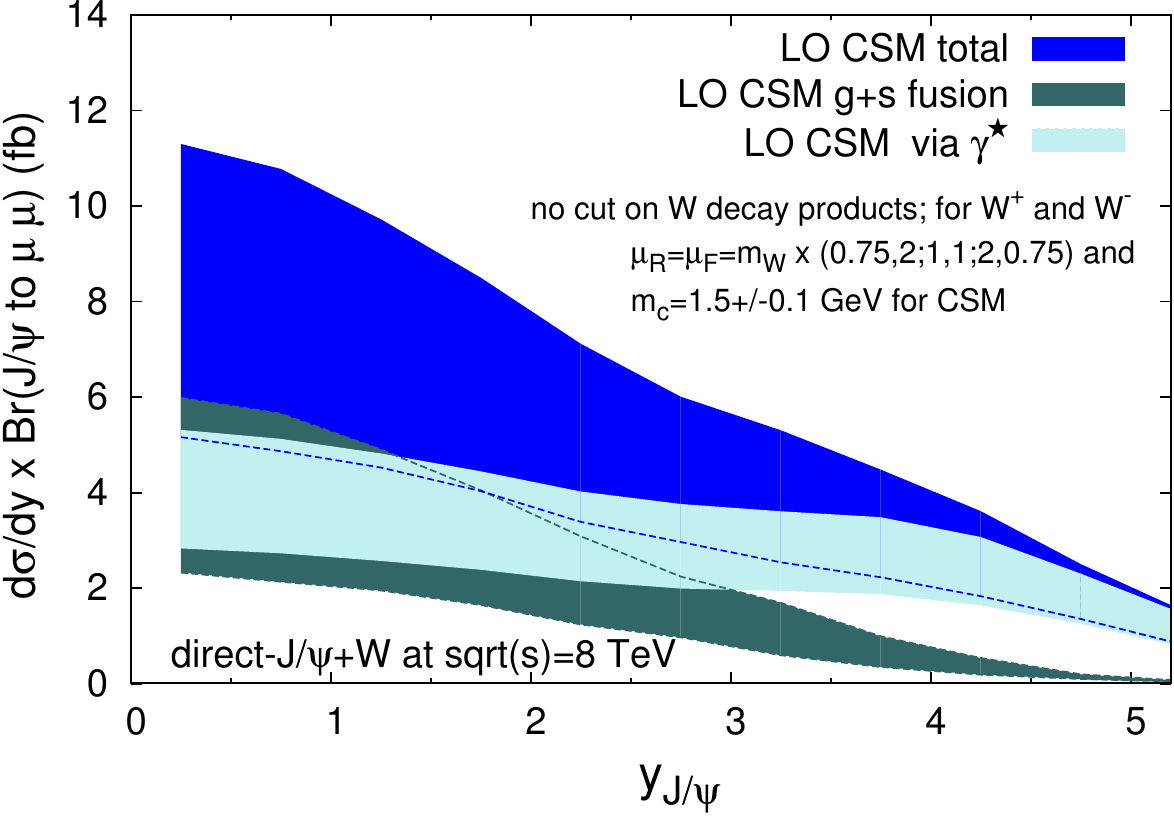}}\\
\subfloat[14 TeV]{\includegraphics[width=.8\columnwidth,draft=false]{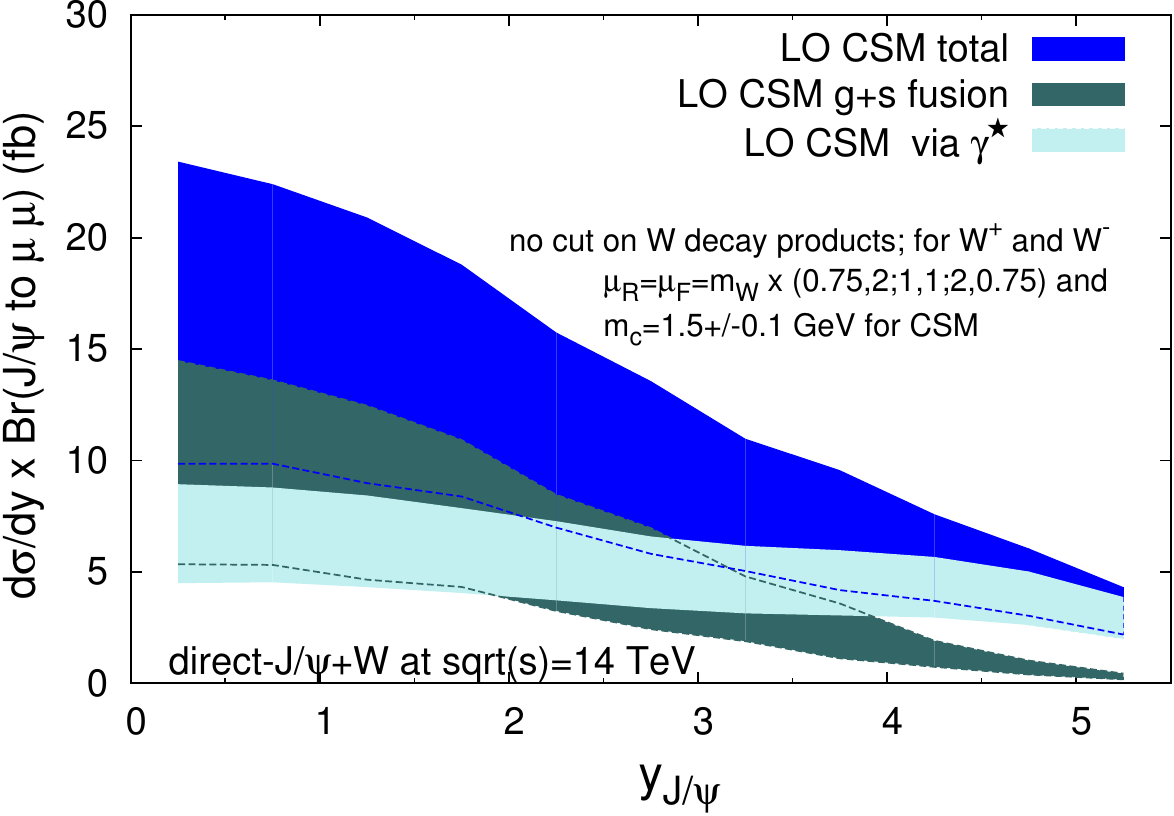}}
\caption{Differential cross section at LO for $J/\psi+W$ vs. $y$ for the LHC at $8$ TeV $(a)$ and $14$ TeV $(b)$. The colour code is the same as in figure \ref{fig:dsigdPT}. Note that these results are obtained without cut on the $J/\psi$ $P_T$.}
\label{fig:dsigdy}
\end{figure}

In addition to the $P_T$ dependence, we present in \cf{fig:dsigdy}  our CSM results for the differential cross sections in $y$ for the LHC at $8$ TeV $(a)$ and $14$ TeV $(b)$. One observes that the CSM yields via $\gamma^\star$ and via $sg$ fusion are of the same order at the LHC energies, with an increasing proportion of $sg$ fusion as the energy increases.

\section{Singlet contributions via an off-shell photon vs. octet contributions via an off-shell gluon in processes involving quarks}

As we have seen above, the contributions from the CSM via an off-shell photon and from the COM via an off-shell gluon --namely via a $^3S^{[8]}_1$ state-- are similar, with nearly exactly the same $P_T$ dependence. We have found it instructive to investigate this.  

To this end, we have evaluated the cross section for a $q\bar q$ annihilation into a $^3S_1$ quarkonium in both channels. Apart from the radiation of the $W$, this is the same process as discussed above.

The partonic cross section for the singlet contribution via an off-shell photon, $q(p_1) \bar q(p_2) \to \gamma^\star \to \Q(p_\Q)$, is : 
\begin{equation}
\hat\sigma^{[1]}_{via \ \gamma^\star}=\frac{(4\pi\alpha)^2e_q^2e_Q^2}{M^3_\mathcal Qs}\,\delta\left(x_1x_2-M^2_\mathcal Q/s\right)
|R(0)|^2,
\end{equation}
with $\hat s=(p_1+p_2)^2=s x_1 x_2$, $e_Q$ the heavy quark charge and $e_q$ the light quark charge.

For the octet contribution via a $^3S_1^{[8]}$ state, one can follow Petrelli \emph{et al.}~\cite{Petrelli:1997ge} and obtain 
\begin{equation}
\hat\sigma^{[8]}_{via \ g^\star}=\frac{(4\pi\alpha_S)^2 \pi}{27M^3_\mathcal Qs}\,\delta\left(x_1x_2-M^2_\mathcal Q/s\right)
\langle\mathcal O_{\Q} (\phantom{\!\!\!S}^3S^{[8]}_1)\rangle.
\end{equation}
where $\langle\mathcal O_{\Q} (\phantom{\!\!\!S}^3S^{[8]}_1)\rangle$ is to be fitted to reproduce the $P_T$ spectrum of the data at the Tevatron and at the LHC and, in principle also, the yield polarisation. In the singlet case, one can connect the wave function at the origin to a similar matrix element: $\langle O_{\Q}(^3S_1^{[1]})\rangle=2N_c(2J+1)\frac{|R(0)|^2}{4\pi}$.

One can now make the ratio of the singlet to octet contribution: 
\begin{equation}
\frac{\hat\sigma^{[1]}_{via \ \gamma^\star}}{\hat\sigma^{[8]}_{via \ g^\star}}=\frac{6\alpha^2e_q^2e_Q^2\langle
\mathcal O_{\Q} (\phantom{\!\!\!S}^3S^{[1]}_1)\rangle}{\alpha^2_S\langle\mathcal O_{\Q} (\phantom{\!\!\!S}^3S^{[8]}_1)\rangle}.
\end{equation}
The difference in the colour structure gives the relative factor $2N_c$ between the octet and the singlet contributions. In the $J/\psi$ case, we have $\langle\mathcal O_{J/\psi} (\phantom{\!\!\!S}^3S^{[1]}_1)\rangle=1.45$ GeV$^3$, $\langle\mathcal O_{J/\psi} (\phantom{\!\!\!S}^3S^{[8]}_1)\rangle=2.2\times 10^{-3}$ GeV$^3$ as we used above and $\alpha_S(M_{J/\psi})=0.26$. The ratio is then about two thirds for $u\bar u$ fusion. In the case of $\Upsilon$ production ($\langle\mathcal O_{\Upsilon} (\phantom{\!\!\!S}^3S^{[1]}_1)\rangle\simeq 10$~GeV$^3$, $\langle\mathcal O_{\Upsilon} (\phantom{\!\!\!S}^3S_1^{[8]})\rangle=0.4\div 3\,  \times\, 10^{-2}$ GeV$^3$ at LO~\cite{Braaten:2000cm} and $\alpha_S(M_{\Upsilon})=0.16$), the ratio is similar to that of $J/\psi$.

Along the same lines, if the quark line emits a $W$ boson, one expects the same ratio up to factors involving the quark ($q$ and $q'$) electric charges\footnote{In fact, the ratio is expected to become more favourable to the CSM contributions by a factor of 5 for the $J/\psi$ and 2 for the $\Upsilon$, since the natural scale of the process would then be $m_W$ rather than $m_{\cal Q}$; the strong coupling would then be smaller and the electroweak one larger.}. This convincingly explains the similarity between the orange (COM) and light blue (CSM via $\gamma^\star$) bands on \cf{fig:dsigdPT} c) for instance. 

Both observations can surely be extended to NLO in $\alpha_S$  and then by using, in a coherent manner, the NRQCD LDME values recently fit in \eg~\cite{Butenschoen:2011yh,Chao:2012iv,Gong:2012ug}. As we mentioned earlier, the LDME values extracted from these recent fits unfortunately depend much on the data sets which were used. From the simple computations done above, it cannot therefore be excluded at all that CSM contributions via an off-shell photon would in fact be larger than that from COM contributions via an off-shell gluons in specific processes, as the ones studied here, where quark--antiquark annihilation is dominant. 

Let us emphasise that the partonic process $q \bar q \to \!\phantom{\!\!\!S}^3S_1$ is expected to take over $gg$ and $gq$ fusions in inclusive $J/\psi$ and $\Upsilon$ production at energies close to threshold. Further quantitative statements however require a dedicated survey within the CSM in particular in what concerns the feed-down from $\chi_c$ 

\section{Additional phenomenological observations}

Beside the discussion of the contributions from CO and CS channels, there are additional important aspects to keep in mind when comparing the theoretical predictions for $J/\psi +W$ to data.

\subsection{$W$ decay channel}

First, experimental analyses of $W$ production usually proceed by looking at the $W$ leptonic-decay products, in particular $\mu+\nu_\mu$. Such events can be tagged by the presence of a missing mass carried by the undetected neutrino. This however also means that one cannot strictly enforce that the invariant mass of the $\mu+\nu_\mu$ pair equals that of the $W$. 

An unexpected consequence of this in the present study is that the (rare) 3-body decay\footnote{A similar decay channel of the $W$, $W\to \Upsilon+\mu+\nu_\mu$ has previously been considered in~\cite{Qiao:2011yk}.} $W\to J/\psi+\mu+\nu_\mu$ cannot be disentangled from genuine $J/\psi+W \to J/\psi+\mu+\nu_\mu$ events. In fact, its contribution is not negligible with the typical cuts used at the LHC. We have indeed found that with the cuts used by ATLAS~\cite{ATLAS:2013kla} ($E^\text{miss}_T >20$ GeV, $P^\mu_T>25$ GeV, $|\eta^\mu|<2.4$, $m^W_T=\sqrt{2P^\mu_TE^\text{miss}_T[1-\cos(\phi^\mu-\phi^\nu)]}>40$ GeV), the process $q\bar q' \to W \to J/\psi+\mu+\nu_\mu$ contributes nearly equally to that of $q\bar q' \to J/\psi+W\to J/\psi+\mu+\nu_\mu$, where $Br(W \to \mu+\nu_\mu)\simeq 11\%$.

\subsection{Double-parton-scattering contributions}

Second, ATLAS has evaluated~\cite{ATLAS:2013kla}, using the pocket formula $\sigma_{DPS}^{J/\psi W}=\sigma^{J/\psi} \sigma^{W}/\sigma_{\rm eff.}$, that a significant Double-Parton-Scattering (DPS) contribution --as high as 40 \%-- is to be expected provided that this formula makes sense and that one can use the effective cross section $\sigma_{\rm eff.}$ as extracted from the $W+2$~jets analysis~\cite{Aad:2013bjm}.

In principle, the DPS signal is reducible since the $J/\psi$ and the $W$ should completely be uncorrelated in $\phi$ and $P_T$. In practice, since one expects only a handful of events per fb$^{-1}$, it will be very complicated to subtract with a good accuracy the SPS signal by looking at the $\Delta\phi$ or $\Delta P_T$ distributions.

\subsection{$\chi_c$ feed-down}

Third, as for most quarkonium-production observables, feed-down from excited-quarkonium states can be important and proceeds from  different partonic reactions. We have indeed computed that the cross section for $\chi_c+W$ times the branching $\chi_c\to J/\psi+\gamma$ is about $6$ times larger than the direct cross section for $J/\psi +W$. In short, the feed-down from $\chi_c$ is expected to be larger than in the inclusive case and cannot be disregarded. This simply comes from the possibility of a fragmentation contribution at $\alpha \alpha_S^3$.

Summing the direct contribution to the feed-down from $\chi_c$ and $\psi(2S)$, we find a total cross-section of $\sigma(|y|<2.4)=4.5\pm 2.3$~fb at $7$ TeV, comparable to the cross-section $\sigma(|y|<2.4)=15\pm 10$~fb for DPS-subtracted prompt $J/\psi+W$ recently obtained  by the ATLAS collaboration \cite{ATLAS:2013kla}.

\section{Conclusions}

We have shown that the LO CSM contributions to direct $J/\psi+W^\pm$  are not negligible compared to the contribution arising from CO transitions which were previously thought to be dominant. These CSM contributions arise from two sub-processes: a) the fusion of a gluon and a strange quark which turns into a charm quark by the emission of the $W$, the charm quark subsequently fragments into a $J/\psi+c$ pair; b) the annihilation of a quark $q$ and an antiquark $\bar q'$ into an off-shell photon, $\gamma^\star$, and a $W$, the $\gamma^\star$ subsequently fluctuates into a $J/\psi$. The former process appears at $\alpha_S^3 \alpha$ and the latter at $\alpha^3$ compared to $\alpha_S^2 \alpha$ for the COM process which is however suppressed in the $v$ expansion of NRQCD.

We have also noted that, for any $^3S_1$ quarkonium-production process involving quark--antiquark annihilation, the CSM process via an off-shell photon numerically competes with the COM one via an off-shell gluon through a $^3S_1^{[8]}$ octet.

Finally, owing to the uncertainties on the CO LDME, the small rate for this process at the LHC and the 
possibility for large DPS contributions, our conclusion is that the study of direct $J/\psi +W$ yields cannot 
serve as a clean probe of the colour-octet mechanism, as previously stated in the literature.


\section*{Acknowledgements}

We thank B. Gong, D. Price and J.X. Wang for useful discussions. This work is supported in part by the France-China Particle Physics Laboratory and by the P2I network.




\begin{thebibliography}{99}


\bibitem{Lansberg:2006dh}
  J.~P.~Lansberg,
  Int.\ J.\ Mod.\ Phys.\  A {\bf 21}, 3857 (2006).

\bibitem{Brambilla:2010cs} 
  N.~Brambilla {\it et al.} 
  Eur.\ Phys.\ J.\ C {\bf 71}, 1534 (2011).

\bibitem{ConesadelValle:2011fw}
  Z.~Conesa del Valle, {\it et al.}, 
  Nucl.\ Phys.\ (PS)  {\bf 214}, 3 (2011).

\bibitem{CSM_hadron}
C-H. Chang,
{Nucl. Phys. } B {\bf 172}, 425 (1980); 
R. Baier and R. R\"uckl,
{Phys. Lett. } B {\bf 102}, 364 (1981); 
R. Baier and R. R\"uckl,
{Z. Phys. } C {\bf 19}, 251 (1983).




\bibitem{Campbell:2007ws}
  J.~Campbell, F.~Maltoni and F.~Tramontano,
  Phys.\ Rev.\ Lett.\  {\bf 98}, 252002 (2007).

\bibitem{Artoisenet:2007xi}
  P.~Artoisenet, J.~P.~Lansberg and F.~Maltoni,
  Phys.\ Lett.\  B {\bf 653}, 60 (2007).

\bibitem{Gong:2008sn}
  B.~Gong and J.~X.~Wang,
  Phys.\ Rev.\ Lett.\  {\bf 100}, 232001 (2008).

\bibitem{Gong:2008hk}
  B.~Gong and J.~X.~Wang,
  Phys.\ Rev.\  D {\bf 78}, 074011 (2008).



\bibitem{Artoisenet:2008fc}
  P.~Artoisenet, J.~Campbell, J.~P.~Lansberg, F.~Maltoni and F.~Tramontano,
  Phys.\ Rev.\ Lett.\  {\bf 101}, 152001 (2008).

\bibitem{Lansberg:2008gk}
  J.~P.~Lansberg,
  Eur.\ Phys.\ J.\  C {\bf 61}, 693 (2009).




\bibitem{Lansberg:2010vq}
  J.~P.~Lansberg,
  Phys.\ Lett.\ B {\bf 695}, 149 (2011).


\bibitem{Li:2008ym}
  R.~Li and J.~X.~Wang,
  Phys.\ Lett.\  B {\bf 672}, 51 (2009).

\bibitem{Lansberg:2009db}
  J.~P.~Lansberg,
  Phys.\ Lett.\  B {\bf 679}, 340 (2009).


\bibitem{Brodsky:2009cf} 
  S.~J.~Brodsky and J.~P.~Lansberg,
  Phys.\ Rev.\ D {\bf 81}, 051502 (2010).

\bibitem{Lansberg:2010cn} 
  J.~P.~Lansberg,
  PoS ICHEP {\bf 2010}, 206 (2010).

\bibitem{Barbieri:1976fp}
  R.~Barbieri, R.~Gatto and E.~Remiddi,
  Phys.\ Lett.\ B {\bf 61} (1976) 465.


\bibitem{Bodwin:1992ye}
  G.~T.~Bodwin, E.~Braaten and G.~P.~Lepage,
  Phys.\ Rev.\ D {\bf 46} (1992) 1914


\bibitem{Barger:1995vx}
  V.~D.~Barger, S.~Fleming and R.~J.~N.~Phillips,
  Phys.\ Lett.\ B {\bf 371}, 111 (1996).

\bibitem{Kniehl:2002wd}
  B.~A.~Kniehl, C.~P.~Palisoc and L.~Zwirner,
  Phys.\ Rev.\ D {\bf 66}, 114002 (2002).


\bibitem{Li:2010hc}
  G.~Li, M.~Song, R.~-Y.~Zhang and W.~-G.~Ma,
  Phys.\ Rev.\ D {\bf 83}, 014001 (2011).

\bibitem{Baur:1993zd}
  U.~Baur, F.~Halzen, S.~Keller, M.~L.~Mangano and K.~Riesselmann,
  Phys.\ Lett.\ B {\bf 318}, 544 (1993).




\bibitem{Artoisenet:2007qm}
  P.~Artoisenet, F.~Maltoni and T.~Stelzer,
  JHEP {\bf 0802}, 102 (2008).

\bibitem{Madonia}
{\small MADONIA}  can be used online (model ``Quarkonium production in SM'') at {\tt \footnotesize http://madgraph.hep.uiuc.edu}.

\bibitem{Shao:2012iz}
  H.~-S.~Shao,
Comput.\ Phys.\ Commun.\  {\bf 184} (2013) 2562

\bibitem{Butenschoen:2011yh}
  M.~Butenschoen and B.~A.~Kniehl,
  Phys.\ Rev.\ D {\bf 84} (2011) 051501

\bibitem{Chao:2012iv}
  K.~-T.~Chao, Y.~-Q.~Ma, H.~-S.~Shao, K.~Wang and Y.~-J.~Zhang,
  Phys.\ Rev.\ Lett.\  {\bf 108} (2012) 242004

\bibitem{Gong:2012ug}
  B.~Gong, L.~-P.~Wan, J.~-X.~Wang and H.~-F.~Zhang,
  Phys.\  Rev.\  Lett {\bf 110}, 042002 (2013)


\bibitem{Braaten:1999qk}
  E.~Braaten, B.~A.~Kniehl and J.~Lee,
  Phys.\ Rev.\ D {\bf 62} (2000) 094005






\bibitem{Kniehl:1998qy}
  B.~A.~Kniehl and G.~Kramer,
  Eur.\ Phys.\ J.\ C {\bf 6} (1999) 493

\bibitem{Gong:2008ft}
  B.~Gong, X.~Q.~Li and J.~-X.~Wang,
  Phys.\ Lett.\ B {\bf 673} (2009) 197
   [Erratum-ibid.\  {\bf 693} (2010) 612]



\bibitem{Petrelli:1997ge} 
  A.~Petrelli, M.~Cacciari, M.~Greco, F.~Maltoni and M.~L.~Mangano,
  Nucl.\ Phys.\ B {\bf 514}, 245 (1998).




\bibitem{Braaten:2000cm}
  E.~Braaten, S.~Fleming and A.~K.~Leibovich,
  Phys.\ Rev.\ D {\bf 63} (2001) 094006

\bibitem{Qiao:2011yk}
  C.~-F.~Qiao, L.~-P.~Sun, D.~-S.~Yang and R.~-L.~Zhu,
  Eur.\ Phys.\ J.\ C {\bf 71} (2011) 1766

\bibitem{ATLAS:2013kla}
  [ATLAS Collaboration],
  ATLAS-CONF-2013-042.

\bibitem{Aad:2013bjm}
  G.~Aad {\it et al.}  [ATLAS Collaboration],
  New J.\ Phys.\  {\bf 15} (2013) 033038


\expandafter\ifx\csname natexlab\endcsname\relax\def\natexlab#1{#1}\fi
\expandafter\ifx\csname bibnamefont\endcsname\relax
  \def\bibnamefont#1{#1}\fi
\expandafter\ifx\csname bibfnamefont\endcsname\relax
  \def\bibfnamefont#1{#1}\fi
\expandafter\ifx\csname citenamefont\endcsname\relax
  \def\citenamefont#1{#1}\fi
\expandafter\ifx\csname url\endcsname\relax
  \def\url#1{\texttt{#1}}\fi
\expandafter\ifx\csname urlprefix\endcsname\relax\def\urlprefix{URL }\fi
\providecommand{\bibinfo}[2]{#2}
\providecommand{\eprint}[2][]{\url{#2}}

\end{thebibliography}
\end{document}